\documentclass{iau}
\usepackage{graphicx}

\title[Magnetic fields in young solar-type stars] 
{The evolution of surface magnetic fields in young solar-type stars}

\author[C.P. Folsom, P. Petit, J. Bouvier, J. Morin \& A. Lebre]   
{Colin P. Folsom$^{1,2}$,
  Pascal Petit$^{3}$,
  J\'er\^ome Bouvier$^{1,2}$,
  Julien Morin$^4$, \\
  Agn\`es L\`ebre$^4$,
  \and Jean-Fran\c cois Donati$^{3}$
}

\affiliation{$^1$Universit\'e Grenoble Alpes, IPAG, F-38000 Grenoble, France\\
$^2$CNRS, IPAG, F-38000 Grenoble, France\\email: {\tt colin.folsom@obs.ujf-grenoble.fr}\\
$^3$IRAP, CNRS and Universit\'e de Toulouse, 14 avenue \'Edouard Belin, 31400, Toulouse, France\\
$^4$LUPM, UMR 5299, CNRS and Universit\'e Montpellier II - Place E. Bataillon, 34090 Montpellier, France\\
}

\pubyear{2015}
\volume{314}  
\pagerange{1--4}
\jname{Young Stars \& Planets Near the Sun}
\editors{J. H. Kastner, B. Stelzer, \& S. A. Metchev, eds.}
\begin{document}

\maketitle

\begin{abstract}
Surface rotation rates of young solar-type stars display drastic changes at the end of the pre-main sequence through the early main sequence. This may trigger corresponding changes in the magnetic dynamos operating in these stars, which ought to be observable in their surface magnetic fields. We present here the first results of an observational effort aimed at characterizing the evolution of stellar magnetic fields through this critical phase. We observed stars from open clusters and associations, which range from 20 to 600 Myr, and used Zeeman Doppler Imaging to characterize their complex magnetic fields. We find a clear trend towards weaker magnetic fields for older ages, as well as a tight correlation between magnetic field strength and Rossby number over this age range. Comparing to results for younger T Tauri stars, we observe a very significant change in magnetic strength and geometry, as the radiative core develops during the late pre-main sequence. 
\end{abstract}

\firstsection 
\section{Introduction}

Solar type stars display a major change in their rotation rates as they cross the pre-main sequence (PMS) and settle on to the main sequence (e.g.,\ \cite[{Gallet} \& {Bouvier} 2013]{Gallet2013-Bouvier-ang-mom-evol}, \cite[2015]{Gallet2015-Bouvier-ang-mom-evol2}).  Early on the PMS, stars strongly interact with their disks, and this regulates their rotation rates.  After a few Myr, a star decouples from its disk and is still contracting, which results in a rapid spin up. On a longer timescale, the star loses angular momentum through its magnetized wind, and starts to spin down as it settles onto the main sequence.  

Magnetic fields in solar-type stars are thought to be generated by a dynamo process, possibly an $\alpha$-$\Omega$ dynamo. Thus, the large change in rotation rates around the zero-age main sequence (ZAMS) should have a large impact on magnetic properties. Conversely, the rotational spin down is controlled by magnetic fields (e.g.\ \cite[Matt et~al.\ 2012]{Matt2012-magnetic-breaking-formulation}). It is thus crucial to understand the  connection between magnetism, rotation, and age in order to to fully decipher early stellar evolution and its potential impact on, e.g., the early evolution of planetary systems and their habitability.  

We present here the first results from an investigation of the magnetic properties of young solar-type stars as a function of rotation rate and age.  

\section{Observations}
Targets for our study are members of known moving groups and clusters, in the age range of $\sim$20 to $\sim$600 Myr, when most of the rotational evolution occurs (cf. \cite[Bouvier et~al.\ 2014]{Bouvier2014-rot-evol-review}).  We exclude T Tauri stars, to avoid stars whose rotation rate is still regulated by disk interactions, and that have been studied in previous projects (e.g.\ \cite[Donati et~al.\ 2011]{Donati2011-V4046Sgr-ZDI}).  We also exclude older stars from our selection, which have converged onto the gyrochronologic sequence and whose magnetic properties have been already characterized (e.g. \cite[Petit et~al.\ 2008]{Petit2008-toroiral-poloidal}).  Indeed, our sample was devised to fill the evolutionary gap between early PMS stars and mature main sequence stars. We focus on 0.7-1.0$M_\odot$ stars with known rotational periods, usually derived from photometric monitoring.  

In order to characterize their magnetic fields, the targets were observed with high-resolution spectropolarimetry.  We used ESPaDOnS at the Canada France Hawaii Telescope and Narval at the T\'elescope Bernard Lyot in France.  These are essentially identical spectropolarimeters, with a resolution of 65000, and a wavelength coverage of 3700 to 10500 \AA.  Reduced observations contain a circularly polarized Stokes $V$ spectrum as well as a total intensity Stokes $I$ spectrum (see \cite[Donati et~al.\ 1997]{Donati1997-major} for data reduction).  Observations of a target consist of a series of typically 15 observations, spread over the rotation cycle of the star.  This phase-resolved series of observations allows us to both diagnose the presence of a magnetic field, and characterize its geometry, as discussed below.  

\section{Analysis}

The high resolution, high S/N spectra allow us to derive precise fundamental parameters for the stars.  Using spectrum synthesis, and by fitting model spectra directly to the observations, we derive $T_{\rm eff}$, $\log g$, $v\sin i$, microturbulence, and [Fe/H] for all the stars in our sample.  The spectra can also be used to determine detailed chemical abundances (e.g. Lithium).  Most of the stars have reliable distance estimates, and thus we can determine intrinsic luminosities, and from that radii and masses.  

Accurate rotation periods are essential for this analysis.  Thus we re-derived the rotation periods for the stars in this study.  For most stars, our results support the literature photometric periods.  However, for a few stars, the literature periods are inconsistent with our spectropolarimetric time series, and for those stars we use the periods we derive. 

\begin{figure}[htb]
  \centering
  \includegraphics[width=1.3in]{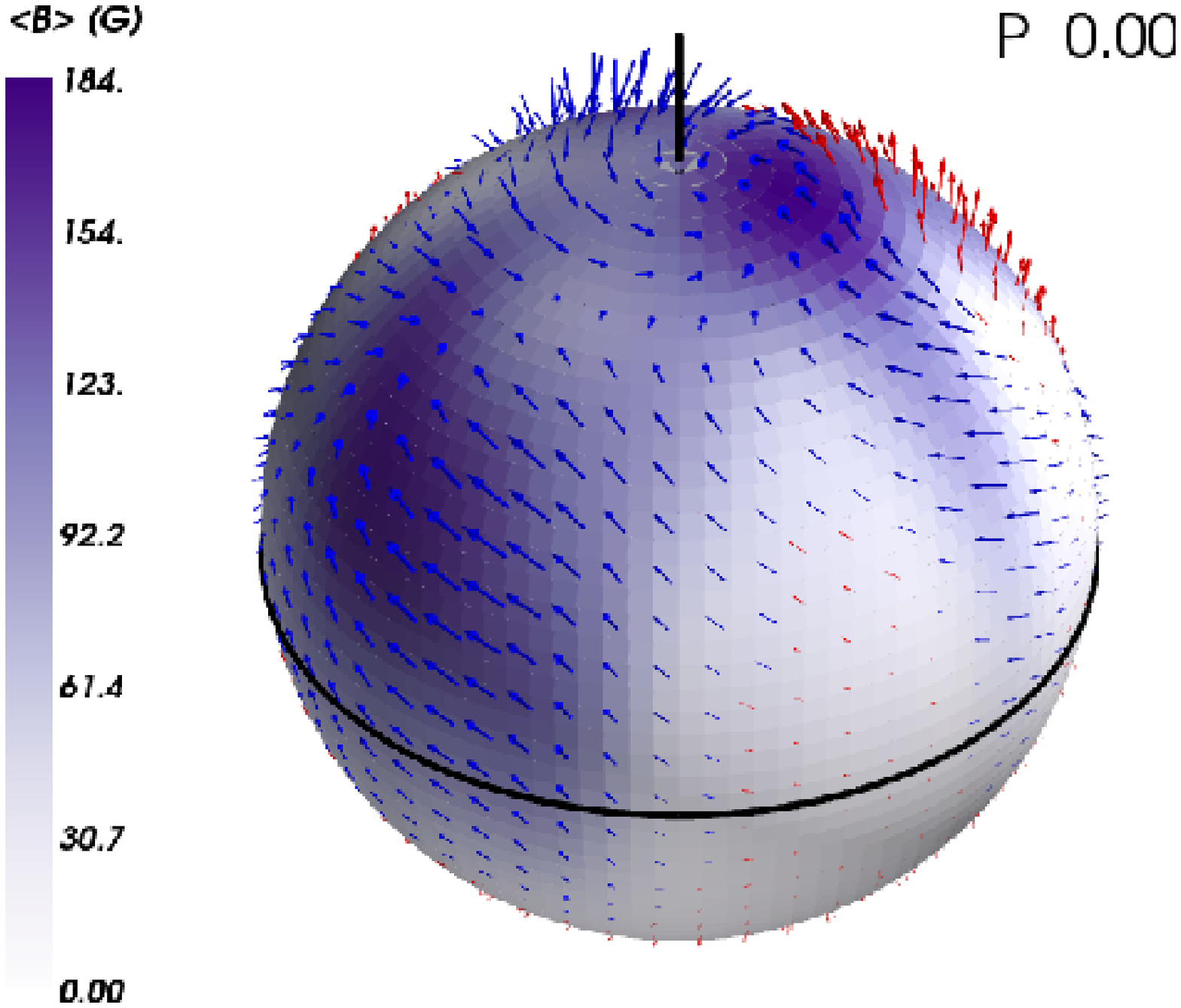} 
  \includegraphics[width=1.3in]{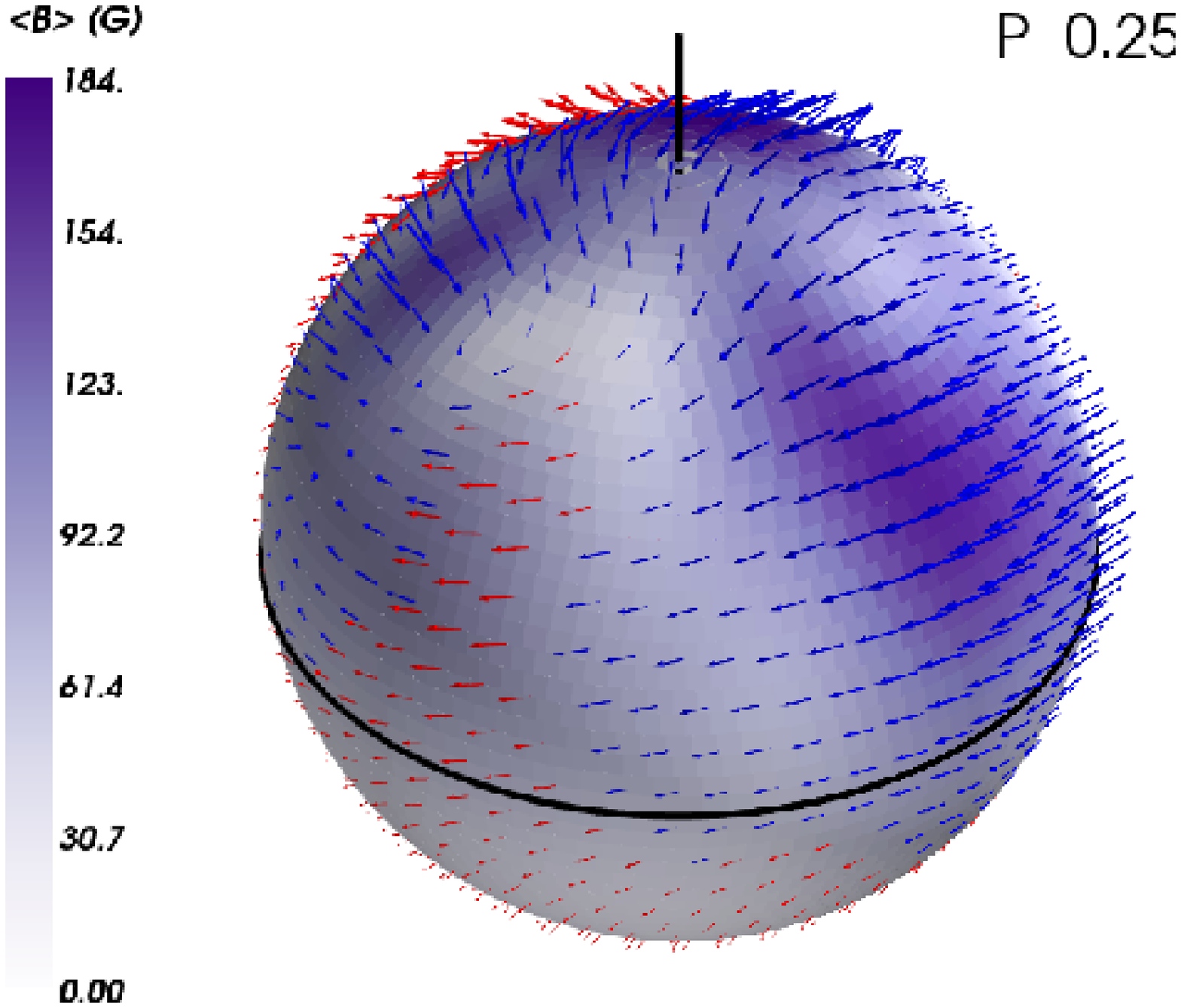}
  \includegraphics[width=1.3in]{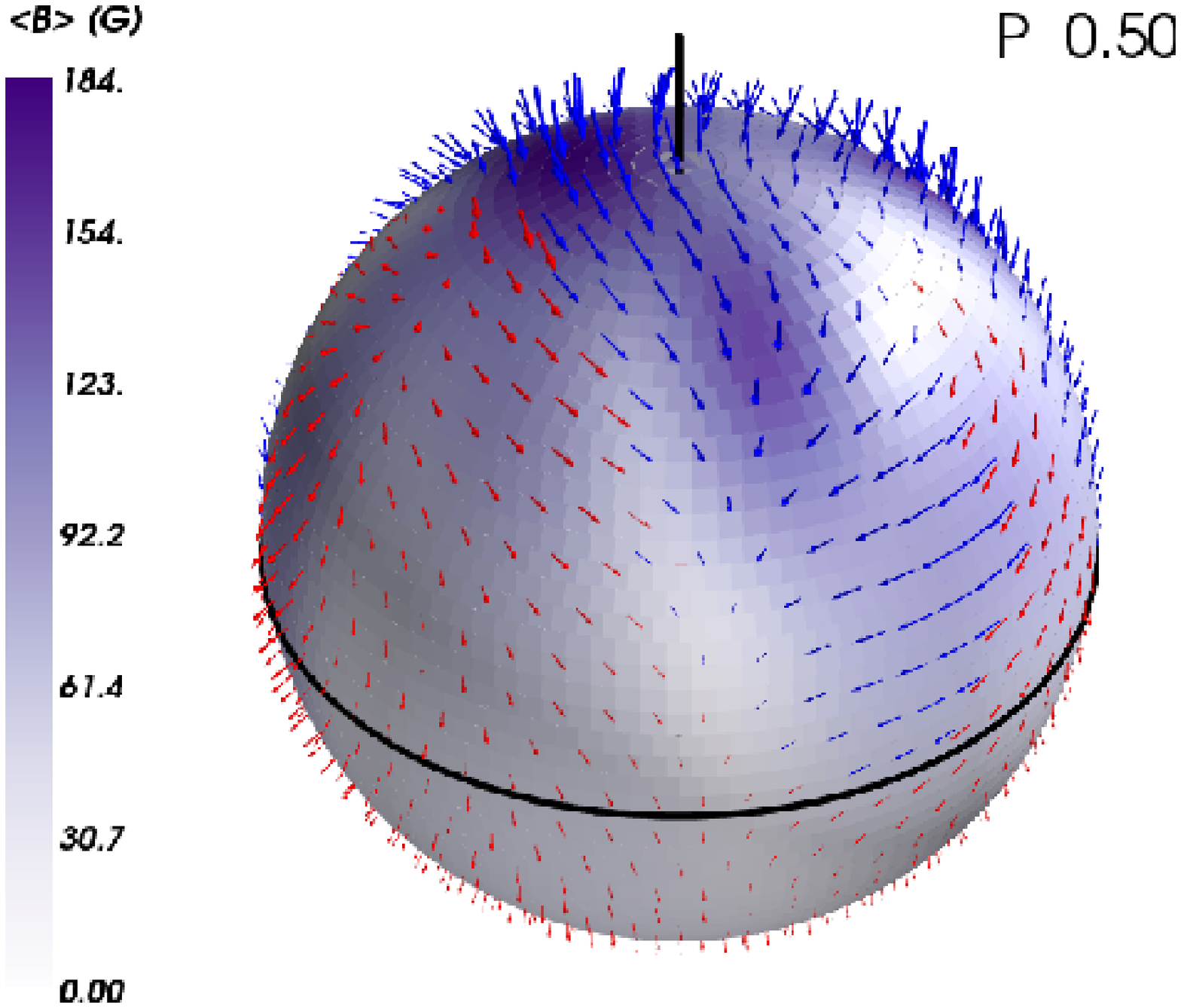}
  \includegraphics[width=1.3in]{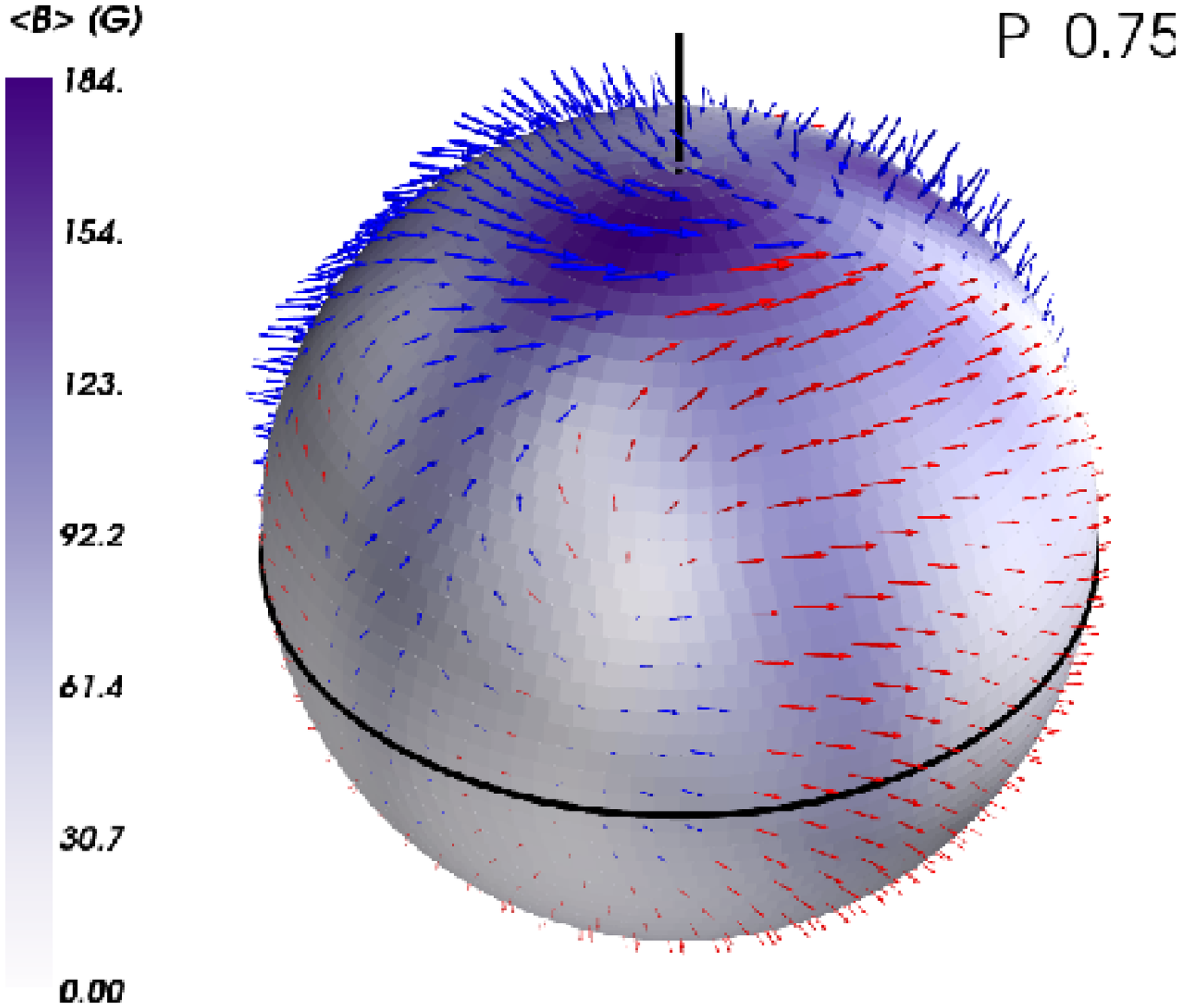}
  \caption{Example of a magnetic map for TYC 6349-0200-1 produced using ZDI.  Arrows indicate magnetic field orientation, 
red arrows have a positive radial component, and blue arrows have a negative radial component. }
  \label{zdi-map-ex}
\end{figure}

In order to derive the magnetic field strengths and geometries of the stars in this study, we use the Zeeman Doppler Imaging (ZDI; \cite[Donati et~al.\ 2006]{Donati2006-tauSco}).  ZDI is a tomographic technique. The rotationally modulated variability of the line profile is inverted to reconstruct the stellar magnetic field.  Stokes $V$ line profiles generated by Zeeman splitting are used.  Our version of the method uses a spherical harmonic decomposition to describe the magnetic field, and the inversion process is regularized using a maximum entropy regularization in order to provide a stable unique solution.  The $v\sin i$ and rotation periods derived above were used as input, as well as an inclination of the rotation axis based on $v\sin i$, radius, and period.  An example magnetic map from ZDI is presented in Fig. \ref{zdi-map-ex}.

\section{Discussion}

Even though our survey of magnetic field properties in young stars is not fully completed yet, some early trends are apparent.  We observe a significant decrease of the global average magnetic field strength with age, as shown in Fig.~2. 
The trend is already seen at the ZAMS, albeit with a large dispersion at a given age, and considerable overlap between the mean magnetic field distributions of stars over the age range from 20 to 250 Myr. The same trend extends all the way to the MS, as previously reported by Vidotto et al. (2014) using the Bcool sample (Petit et al., in prep.).  We find an even tighter correlation between average magnetic strength and Rossby number ($R_o$), as shown in Fig.~2. Indeed, our ZAMS sample complements and extrapolates to younger ages the $\langle B\rangle$-$R_o$ correlation seen on the MS for the Bcool sample. Only the fastest rotator of our sample, LO Peg, may show signs of magnetic saturation at very low Rossby number. 

An interesting comparison sample are the classical T Tauri stars (cTTS) from the MaPP project (e.g.\ \cite[Donati et al.~2011]{Donati2011-V4046Sgr-ZDI}).  These are younger pre-main sequence stars that are still actively accreting and are predominantly convective.  We see a clear distinction between the magnetic properties of the MaPP sample and our older stars.  cTTS have much stronger magnetic fields, and their fields are much more poloidal and aligned with the stellar rotation axis (Fig.~\ref{confusoG}).  This dramatic difference likely reflects a change in the internal structure of the stars, going from fully convective at the T Tauri stage to being largely radiative on the ZAMS. This structural change appears to strongly affect the type of dynamo operating in PMS stars, as suggested by \cite{Gregory2012-TTauri-mag-struct}.  A similar difference is seen between partially and full convective M dwarfs (\cite[Morin et al.~2010]{Morin2010-Mdwarf-mag}).  

The three samples are shown together in Fig.~3, where Rossby number is plotted as a function of age.  It illustrates two main trends: i) the evolution of magnetic properties from the early PMS to the end of the PMS appears to be mainly driven by structural changes, with significant differences seen in magnetic strengths and topologies of PMS and ZAMS stars at a given Rossby number, while ii) the evolution of magnetic properties from the late-PMS to the ZAMS and MS strongly correlates with Rossby number, presumably reflecting the decreasing efficiency of the stellar dynamo as the  stars are spun down. Beyond these clear trends, it is worth noticing the large scatter of magnetic properties seen on the ZAMS. While intrinsic variability contributes to this scatter, it could be related to a third parameter, possibly the large amount of internal differential rotation predicted at this phase of evolution (Gallet \& Bouvier 2013, 2015).

\begin{figure}[htb]
  \centering
  \includegraphics[width=2.6in]{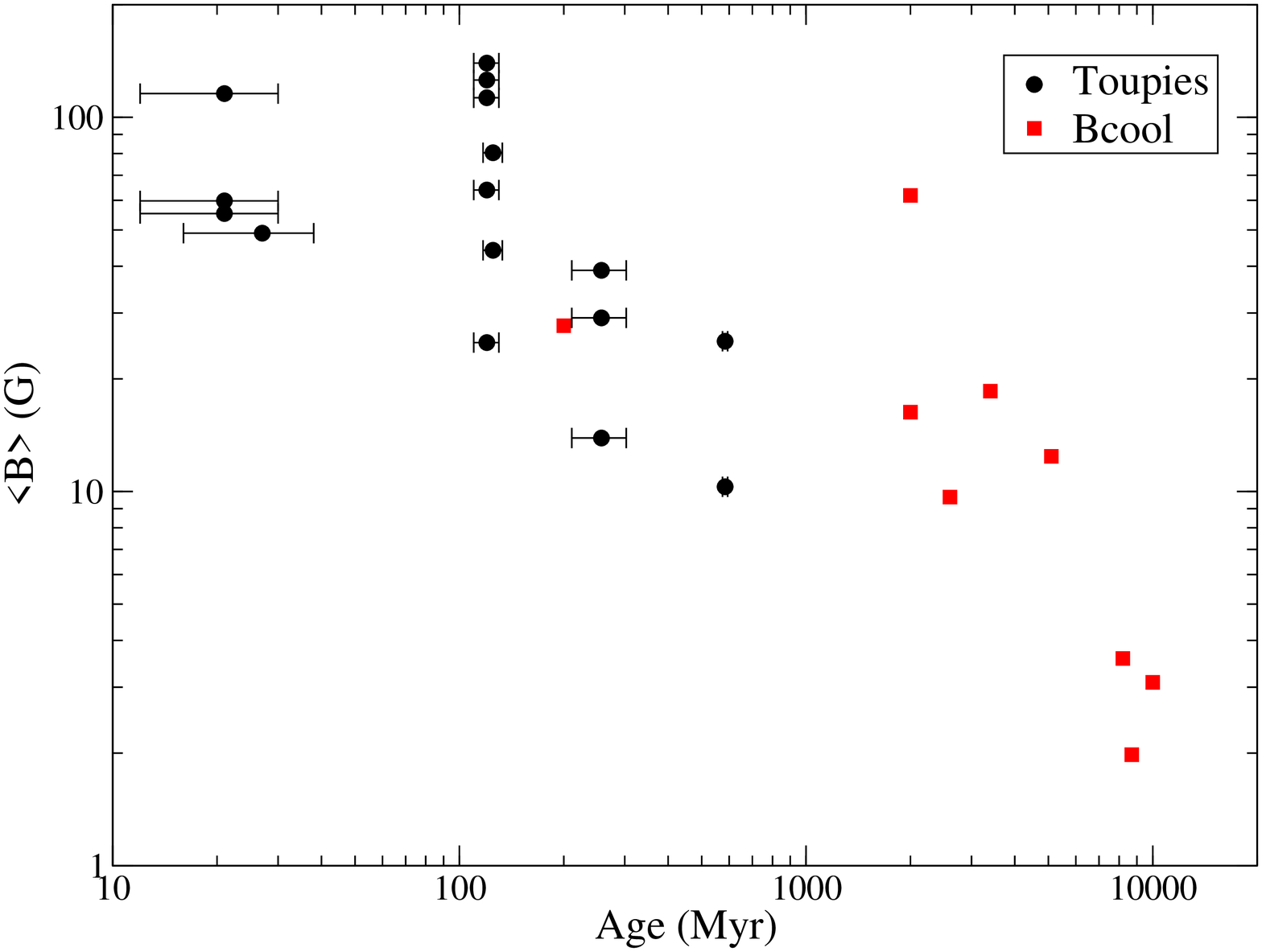} 
  \includegraphics[width=2.6in]{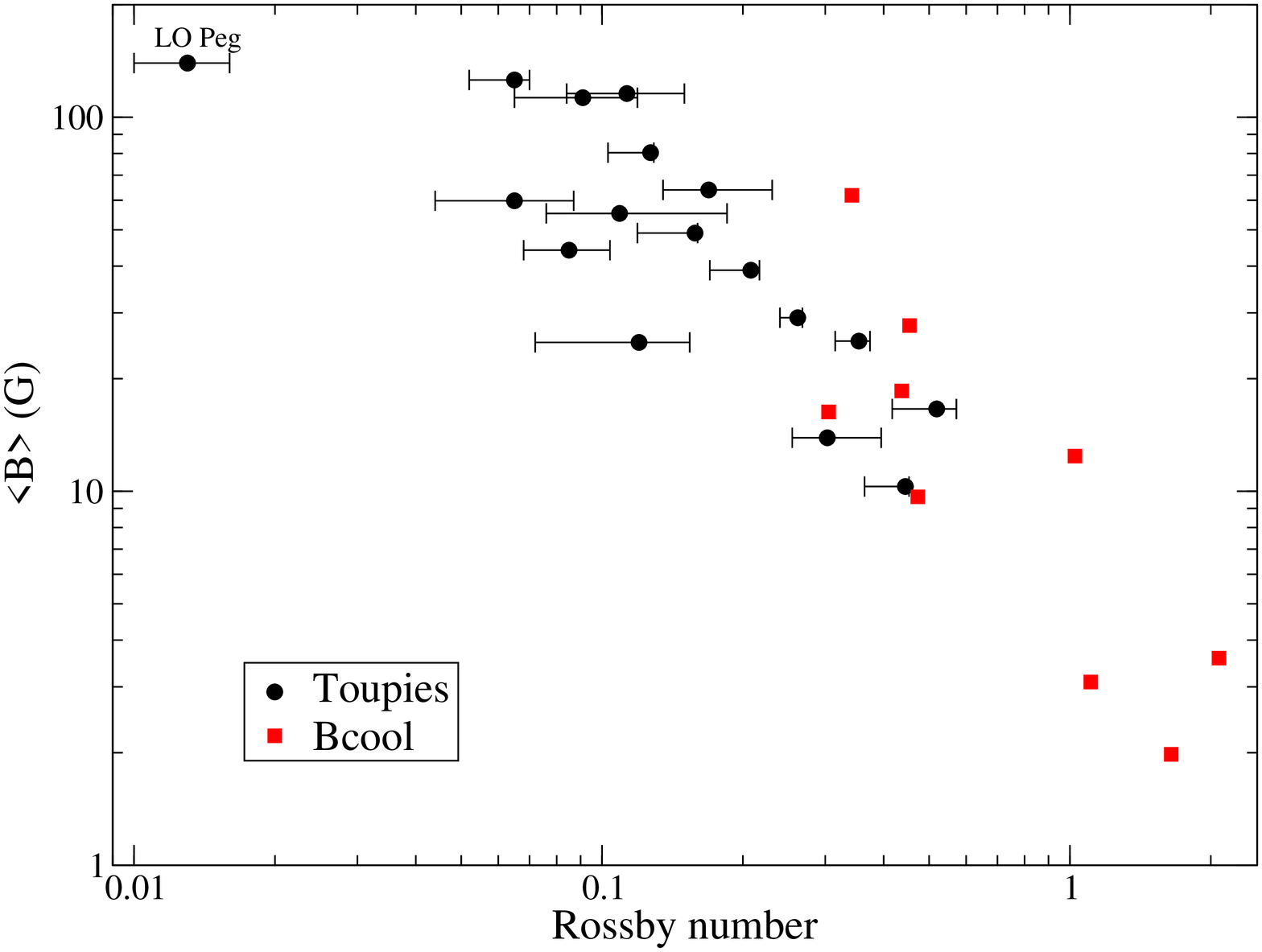}
  \caption{Trends in magnetic field strength with age (left) and Rossby number (right).  Black points are from our sample, and red points are from the Bcool sample of Petit et al.\ (in prep.).  A power law fit of the Rossby number trend produces $\langle B\rangle \propto R_{o}^{-1.0 \pm 0.1}$.  }
  \label{trends-B}
\end{figure}

\begin{figure}[htb]
  \centering
  \includegraphics[width=2.6in]{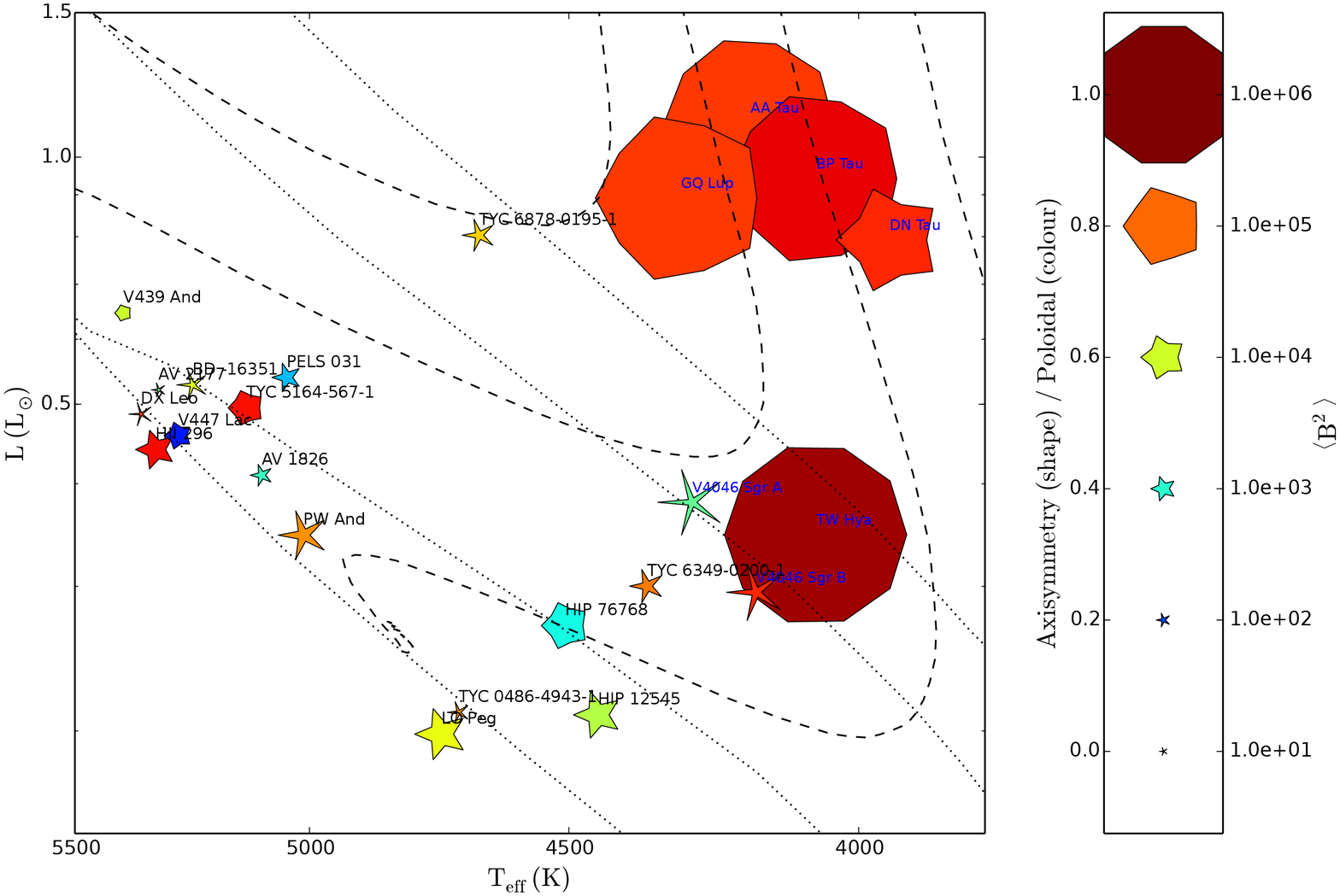} 
  \includegraphics[width=2.6in]{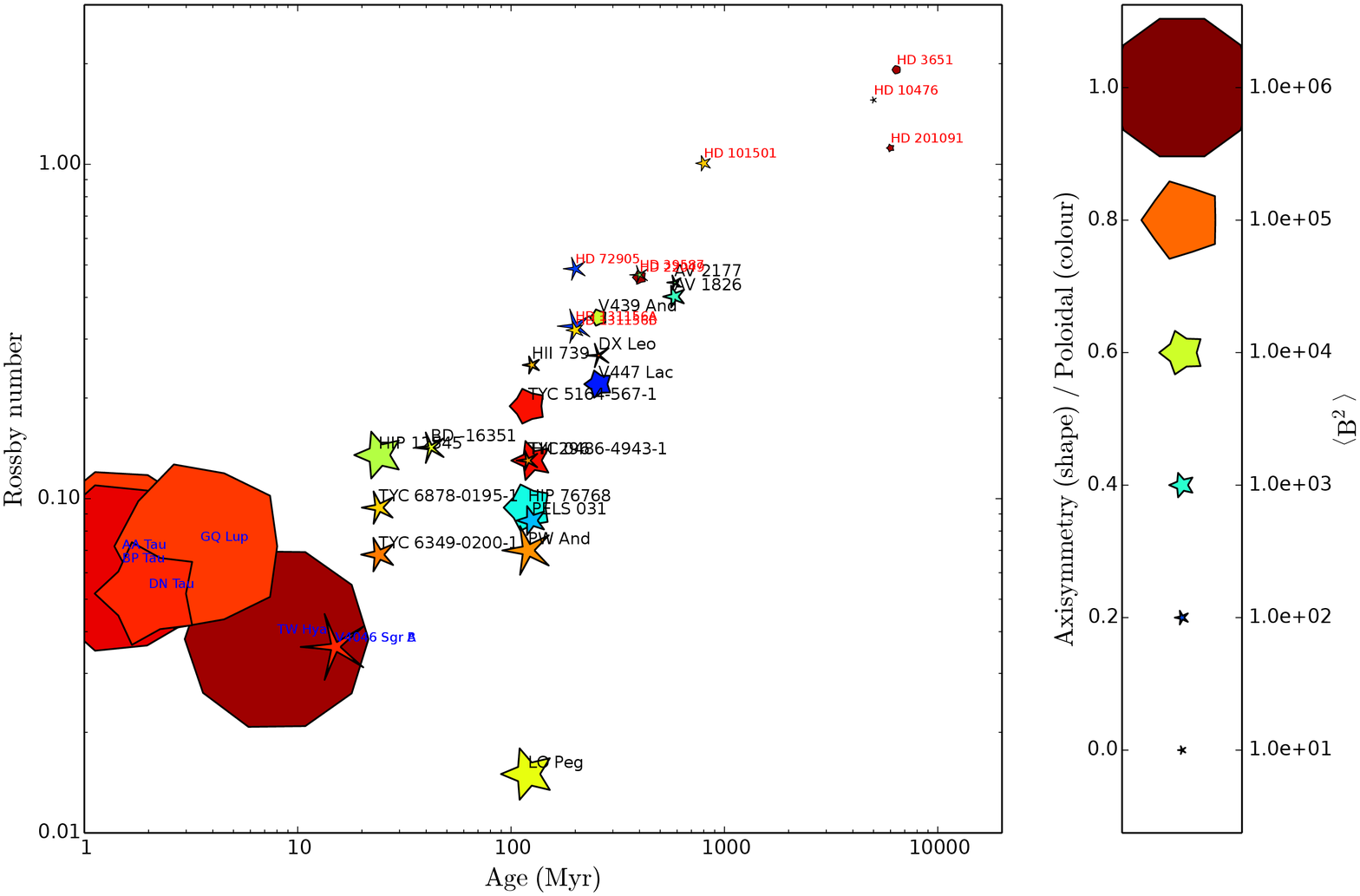}
  \caption{Magnetic properties of stars in our sample (black labels), stars in the Bcool sample (red labels), and stars in the MaPP sample (blue labels).  Left frame: our stars and the MaPP sample in an HR diagram, with pre-main sequence evolutionary tracks.  Right frame: our stars, the Bcool stars, and the MaPP stars in the age-Rossby number plane.  Symbol size corresponds to magnetic field strength, symbol color corresponds to the ratio of poloidal to toroidal field, and shape corresponds to the degree of axisymmetry of the field. }
  \label{confusoG}
\end{figure}

\section{Conclusion}

By studying the magnetic properties of a sample of young solar-type stars, we have filled the evolutionary gap between young PMS T Tauri stars and mature main sequence stars. Comparing samples, the main finding is that magnetic properties scale primarily with internal structure during PMS evolution, and with rotation during ZAMS/early-MS evolution. The transition from a fully convective to a partly radiative interior during the PMS yields complex non-axisymmetric fields on the ZAMS, while the spin down of stars on the MS drives the decline of magnetic field strength on a longer timescale.

While we have drawn some preliminary conclusions here, the study is ongoing with a Large Program being performed at the CFHT (`The History of the Magnetic Sun', PI P. Petit).  This should help us further investigate differences in magnetic properties with rotation at a specific age, and differences with age at a specific Rossby number.

\end{document}